\documentstyle[aps,epsf,floats,
	       twocolumn,
              ] {revtex}

\begin{document}

\title  {       A Monte-Carlo study of meanders
}
\author {       O. Golinelli
}
\address{       Cea Saclay, Service de physique th\'eorique,
\\              91191 Gif-sur-Yvette, France
\\              email: golinelli@cea.fr
%               fax: (+33) 1 69 08 81 20
}
\date   {       June 1999, Sept.\ 1999
}
\preprint{t99/045 ; cond-mat/9906329}
\maketitle

%%%%%%%%%%%%%%%%
\begin{abstract}
%%%%%%%%%%%%%%%%

We study the statistics of meanders, i.e. configurations of a road
crossing a river through $n$ bridges, and possibly winding around the
source, as a toy model for compact folding of polymers.  We introduce a
Monte-Carlo method which allows us to simulate large meanders up to
$n=400$.  By performing large $n$ extrapolations, we give asymptotic
estimates of the connectivity per bridge $R=3.5018(3)$, the
configuration exponent $\gamma=2.056(10)$, the winding exponent $\nu =
0.518(2)$ and other quantities describing the shape of meanders.

\noindent Keywords : folding, meanders, Monte-Carlo, tree

\end{abstract}

\draft
\pacs {PACS number:
	64.60.-i,	% phases transitions
	05.10.Ln,	% Monte-Carlo
	02.10.Eb	% combinatorics
}
\narrowtext

%%%%%%%%%%%%%%%%%%%%%%
\section{Introduction}
%%%%%%%%%%%%%%%%%%%%%%

The concept of folding has an important place in polymer
physics~\cite{cj,gop}.  Considering a polymer chain made of $n$ identical
constituents (the {\em monomers}), the entropy of such a system can be
obtained by counting the number of inequivalent ways of folding the chain
onto itself.  If the model of polymer does not take self-avoidance into
account, it is then equivalent to the well-known Brownian motion.  Several
more involved models have been proposed which study the {\em compact}
folding of a {\em self-avoiding} polymer as a Hamiltonian cycle (i.e. a
closed, self-avoiding walk which visits each vertex) on a regular
lattice\cite{bgo1,bgo2}.  They can also be defined for some kinds of {\em
random} lattices~\cite{egk,gkn}, where each configuration is now described
by a system of non-intersecting arches which connect the pairs of monomers,
which are neighbors in the real space.

In the present paper, the compact folding of a polymer chain is modeled by
a folded strip of stamps, with a complete piling of the strip on top of one
stamp~\cite{saintelague}.  It is then equivalent to another model of
non-intersecting arches, the so-called {\em meander} problem, which can be
summarized by this simple question: in how many ways $M_n$ can a road cross
a river through $n$ bridges, and possibly wind around the source.

A related problem can be defined by forbidding the winding around the source
(i.e.  the river is infinite at the both ends).  It is now equivalent to
enumerate the ``simple alternating transit mazes''~\cite{phillips} of depth
$n$; it was also investigated in connection with Hilbert 16'th problem,
namely the enumeration of ovals of planar algebraic curves~\cite{arnold}.

By analogy with some models of statistical mechanics like random walks or
self-avoiding walks, the meanders can be described in the language of
critical phenomena.  In particular, the asymptotic behavior of meanders when
$n$ is large can be characterized by ``critical'' exponents.  However the
exact enumeration of meanders is particularly complicated : there is no
known formula for $M_n$ in terms of $n$.  By generating all possible
configurations, by hand or with a computer, the beginning of the sequence
$M_n$ can be computed~\cite{touchard,lunnon,koehler,dfgg1,dfgg3,sloane}
exactly.  As $M_n$ increases exponentially with $n$, the limits of computers
are reached for $n\sim 30$ and the estimates of the exponents are too
inaccurate to validate (or invalidate) some conjectures.

One should mention that several exact results, for arbitrarily large $n$,
which are unfortunately not helpful to determine the values of the
exponents, have been obtained with other techniques: random matrix model
methods~\cite{dfgg1,lando_zvonkin,confit,ck} and an algebraic approach using
the Temperley-Lieb algebra~\cite{dfgg2,dif64} or the Hecke
algebra~\cite{dif65}.

Many models in statistical physics can be studied by Monte-Carlo (or
stochastic) methods.  With these algorithms, only a small set of
configurations among all the possible ones are generated.  In principle, the
expectation of physical quantities (like energy or magnetization) with a
given law of probability (like the Boltzmann law involving an external
temperature) can be estimated from these randomly generated samples, if
their probabilities of generation are known.  For example, with the
Metropolis algorithm for classical spin systems~\cite{binder5b}, the
probability of generation is built to be equal to the Boltzmann law and the
average is done over the generated configurations with equal weights.  To
bypass some difficulties (for example when the phase space has many local
minima with high free energy barriers between them), it is possible, in
principal, to generate the random configurations with another more adapted
law and to correct this bias when the average is done~\cite{bnb}.

But for the meanders problem, the situation is quite different~: the phase
space is not easy to built because the number $M_n$ of configurations is
unknown and the only known efficient method to draw a meander of size $n$ is
a recurrence over $n$.  Moreover the na\"{\i}ve way to use this recurrence
gives a distribution of meanders which is not flat, and this default
increases exponentially with $n$.  This paper presents a Monte-Carlo method
which explores the meanders with an almost flat distribution law.
Furthermore the bias is known and can be corrected exactly.  Therefore, the
average can be done over meanders with equal probabilities.  In particular,
better estimates of critical exponents are obtained.

After Section 2 devoted to the definitions, and Section 3 which explains the
building of meanders by recurrence, Section 4 of this paper describes the
Monte-Carlo method.  The results are presented and discussed in Section 5.

%%%%%%%%%%%%%%%%%%%%%
\section{Definitions}
%%%%%%%%%%%%%%%%%%%%%

A {\sl meander} of size $n$ is a planar configuration of a
non-self-intersecting loop ({\sl road}) crossing a half line
(semi-infinite {\sl river} with a {\sl source}) through $n$ points
({\sl bridges}).  Two meanders are considered as {\sl equivalent} if
their roads can be continuously deformed into each other, keeping the
bridges fixed: this is therefore a topological equivalence.  We call an
{\sl arch} each section of road between two consecutive bridges.  So a
meander of size $n$ has $n$ bridges and $n$ arches.

\begin{figure}
  \centering\leavevmode
  \epsfbox{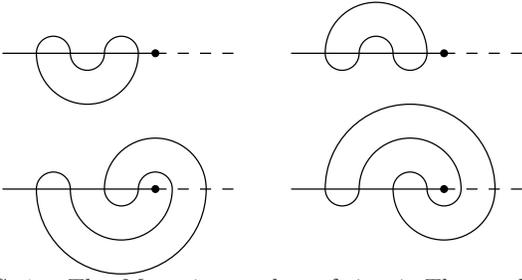}
  \caption{
    The $M_4=4$ meanders of size 4.  The {\sl road} is the
    non-self-intersecting loop. The semi-infinite {\sl river} is the
    solid half-line, starting at the source (black dot).  The {\sl
    size} is the number of bridges.  The {\sl winding} number $w$ is
    the number of arches crossing the dashed half-line on the right
    side.  The upper meanders have no winding ($w=0$), but the lower
    have $w=2$.
  }
  \label{m4}
\end{figure}

The number of different meanders of size $n$ is denoted by $M_n$.  For
example, $M_1 = 1$, $M_2 = 1$, $M_3 = 2$, $M_4 = 4$.  In Fig.~\ref{m4},
the 4 meanders of size 4 are drawn.  The $M_n$'s, up to $n=29$, can be
found in Ref.~\onlinecite{dfgg3,sloane}.

In previous articles~\cite{dfgg1,dfgg3,dfgg2}, these objects were
called {\sl semi}-meanders, to distinguish them from the case where the
line is infinite (river without source).  In this paper, the river is
always a half-line and the word {\sl meanders} is used for convenience.

We can define~\cite{dfgg1,dfgg3,dfgg2} meanders with $k$ {\sl connected
components}, i.e. made of one river and $k$ non-intersecting roads.
But, in this work, we do not include this generalization and we keep
$k=1$.  However the Monte-Carlo method, used in this article, can be
adapted without difficulties to an arbitrary fixed $k$, and even for
varying $k$ with a fugacity $q^k$.

As explained with many details in Ref.~\onlinecite{dfgg3}, the meander
problem is absolutely equivalent to the problem of the compact folding
of a strip of stamps because each meander of size $n$ can be
continuously deformed in such a way that the ``road'' becomes a
vertical line and the ``river'' becomes a folded strip of $n-1$
stamps.  We prefer to present our results with the meander
representation because the main recursion relation, described later,
seems more ``natural'' in this picture.

The meander problem has certain similarities with two-dimensional
self-avoiding walks: a meander is obtained by intersecting a closed
self-avoiding walk by a half-line and keeping only the topological
aspect.  By analogy, it is expected~\cite{lando_zvonkin} that
  \begin{equation}
    M_n \stackrel{n \rightarrow \infty}{\sim} c \; {R^n \over n^\gamma},
    \label{rnng}
  \end{equation}
where the estimates given in Ref.~\onlinecite{dfgg3} are $R=3.50(1)$
and $\gamma \simeq 2$.

The connectivity $R$ can be reinterpreted as the average number of ways
of adding a bridge close to the source by deforming an arch of a given
large meander.  Then $\ln(R)$ is the entropy per bridge.  The
configuration exponent $\gamma$ is sensitive to the boundary
conditions, for example whether the road is closed or open, whether the
river is infinite or semi-infinite, straight or forked, whether the
meander is drawn on a planar surface, a sphere or a surface with higher
genus.  Conversely, we expect that $R$ remains the same for all these
boundary conditions.

It is similar in on-lattice self-avoiding walks problem where the
connectivity depends on the type of lattice (square, honeycomb \dots)
and not on the boundary conditions, whereas the ``universal''
configuration exponent depends on the boundary conditions, but is not
sensitive to the small scale details of the lattice.  For these
reasons, we think that the numerical value of $R$ is valid only for
this particular model of meanders.  But $\gamma$ is expected to be more
``universal'' and to keep its value in other variants of the meander
problem.  Unfortunately, the numerical determination of $\gamma$ is
less precise than $R$, because $n^\gamma$ describes the correction to
the leading exponential asymptotic behavior $R^n$.

For a given meander $m$, the {\sl winding} number $w(m)$ of the road
around the source of the river can be defined as the minimal number of
intersections between the road and a half (semi-infinite) line starting
at the source and extending the river on the opposite side.  For an
example, see Fig.~\ref{m4}.  We define $w_n$ as the average of the
winding number
  \begin{equation}
    w_n = {1 \over M_n} \sum_{m=1}^{M_n} w(m)
    \label{wn}
  \end{equation}
over all the meanders $m$ of size $n$.  We can see the winding number
as the topological end-to-end distance between the source (right end of
the river) and the infinite (left end of the river).  Here the distance
between two points is simply the minimal number of arches which must be
crossed to go from one point to the other.  By analogy with the
end-to-end exponent of self-avoiding walk, we expect that
  \begin{equation}
    w_n \stackrel{n \rightarrow \infty}{\sim}  n^{\nu},
  \end{equation}
where the estimate given in Ref.~\onlinecite{dfgg3} is $\nu=0.52(1)$.

If we study the meanders by leaving free the number $k$ of connected
components, the problem is equivalent~\cite{dfgg3} to a random walk on
a semi-infinite line and can be studied with usual methods of
combinatorics.  In particular, $\gamma = 3/2$, $R=4$ and $\nu = 1/2$ is
the Brownian exponent.  But, by fixing $k=1$, the problem is
drastically more difficult and the above values are, to our knowledge,
not yet known exactly.

\begin{figure}
  \centering\leavevmode
  \epsfbox{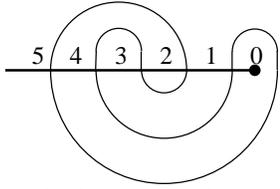}
  \caption{
    The height $h(i)$ (resp. $h(-i)$) is the number of arches over (resp.
    below) the segment $i$.  For this meander of size 5,
    $ \{h(i)\} = \{ 0, 1, 2, 3, 2, 1, 0, 1, 2, 1, 0 \} $
    for $i=-5, \dots 5$.  The area is $A=13$.
  }
  \label{haut}
\end{figure}

For a given meander $m$ of size $n$, we label by $i=0, \dots n$ each
segment of river in-between two consecutive bridges, from right to
left.  So the rightmost segment (with source) is labeled 0, and the
leftmost (semi-infinite) segment is labeled $n$.  We define the {\sl
height} $h(i,m)$ as the number of arches {\sl over} the segment $i$,
and $h(-i,m)$ as the number of arches {\sl below} the segment $i$. An
example is given in Fig.~\ref{haut}.

For the case $i=0$, the both definitions $h(+0,m)$ and $h(-0,m)$ are
equivalent and equal to the winding number: $h(0,m) = w(m)$.  From the
definition, we have $h(n,m)=h(-n,m)=0$, $h(i,m) \geq 0$ and $h(i+1,m)
= h(i,m) \pm 1$.

For a given meander $m$ of size $n$, we define the {\sl area} $A(m)$ as
  \begin{equation}
    A(m) = \sum_{i=-n}^{n} h(i,m).
  \end{equation}
For meanders of size $n$, it can be proved that the maximal area is
$(n-1)^2+1$ for the two meanders with a snail shape (where $ \{h(i)\} =
0, 1, 2, \dots, n-2, n-1, n-2, \dots, 2, 1, 0, 1, 0 \}$ plus the
symmetric meander), and the minimal area is $2n-2$ (resp.  $2n-1$) when
$n$ is even (resp. odd) for the $2^{n/2-1}$ (resp. $2^{(n-1)/2}$) snake
shaped meanders characterized by $h(i)+h(-i) = 2$ for $0<i<n$.  As in
the case of the winding number, we will consider the {\sl average}
profile height
  \begin{equation}
    h_n(i) = {1 \over M_n} \sum_{m=1}^{M_n} h(i,m)
    \label{av.height}
  \end{equation}
and the {\sl average} area
  \begin{equation}
    A_n = {1 \over M_n} \sum_{m=1}^{M_n} A(m)
    \label{av.area}
  \end{equation}
over all the meanders $m$ of size $n$.

%%%%%%%%%%%%%%%%%%%%%%%%%%%%%%%%%%%%%%%%%
\section{Recursion relation for meanders}
%%%%%%%%%%%%%%%%%%%%%%%%%%%%%%%%%%%%%%%%%

In this section, we describe a recursive algorithm to enumerate and
built all meanders of a given size $n$.  Though it was described in
Ref.~\onlinecite{dfgg1,dfgg3}, we prefer to recall it in details,
because the Monte-Carlo method is based on this recursion.

We have different ways to built a meander of size $n+1$, starting from
a meander of size $n$.  Our method consists in adding a bridge on the
left most part of the river (opposite to the source) and changing the
road to cross this new bridge.  To keep this change minimal, only an
{\sl exterior} arch is modified (an arch is {\sl exterior} when no
other arch surrounds it).

\begin{figure}
  \centering\leavevmode
  \epsfbox{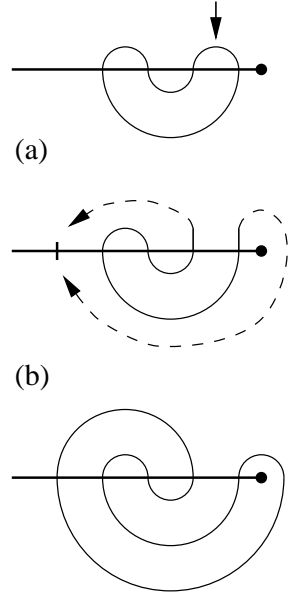}
  \caption{
    A meander of size $n+1$ is built from a meander of size $n$ with
    one labeled exterior arch by the following process. (a) Add a new
    bridge on the left side of the river.  Cut the labeled arch.
    Stretch its two free ends. (b) Close the arch on the opposite side
    by crossing the new bridge (possibly by bypassing the source on the
    right).  The inverse process is the following. (b) Open the road at
    the place of the left most bridge. (a) Pull the two free ends and
    close them on the opposite side to form a {\sl exterior} arch.
  }
  \label{recur}
\end{figure}

Take a meander of size $n$ (the {\sl parent}) and choose (or label) one
of its exterior arch.  By the process described in Fig.~\ref{recur}, a
meander of size $n+1$ (the {\sl child}) is built.  A parent has as many
different children as exterior arches.  By inverting this process, it
appears that each meander of size $n+1$ has one and only one parent.
More precisely, it is a one-to-one mapping between the meanders of size
$n+1$, and the meanders of size $n$ with one labeled exterior arch.

\begin{figure}
  \centering\leavevmode
  \epsfxsize=8.5cm
  \epsfbox{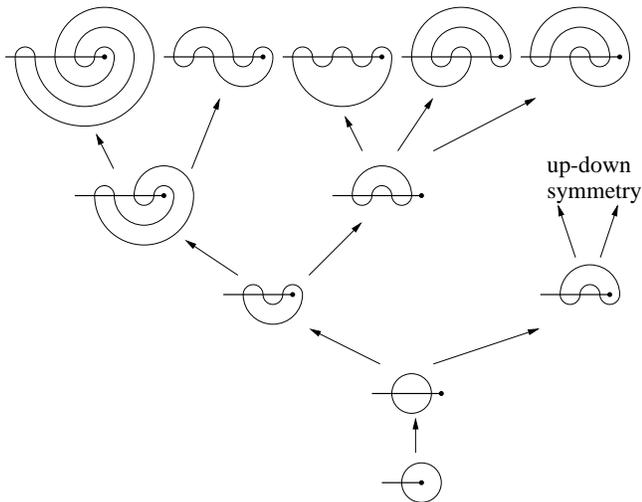}
  \caption{
    The tree of meanders up to $n=5$.  For $n \geq 4$, only a half of
    the tree is drawn by using the up-down symmetry.   Then $M_1 = 1$,
    $M_2 = 1$, $M_3 = 2$, $M_4 = 4$ and $M_5=10$.  Each arrow
    represents a process as described in Fig.~\ref{recur}.
  }
  \label{arbre}
\end{figure}

The starting point of the recursion is the unique meander of size 1.
By $n-1$ successive applications of the recursion process, every
meanders of size $n$ can be built.  As shown in Fig.~\ref{arbre}, the
set of meanders is organized as a {\sl tree}.  The root, at level 1, is
the starting meander $n=1$.  Each branch between a node on level $n$
and a node on level $n+1$ represents a relation between a parent of
size $n$ and its child of size $n+1$.  Apart from $n=1$, a meander (or
node) has several exterior arches, then several children (or
branches).  Their number depends on the precise shape of the parent and
varies between 2 and $n/2+1$.

If we want to exactly enumerate the $M_n$ meanders of size $n$, the
only method we know is to built and investigate the tree up to the
level $n$.  In particular, we have not found a direct recursion between
the numbers $M_n$.  The number of children of each meander has a
distribution which seems to be erratic and the only way to know it is
the examination of its shape.  Then, to compute $M_n$, the work is
proportional to $M_n$.  As the $M_n$'s increase exponentially, the
limits of the capabilities of the computers are rapidly reached.

In Ref.~\onlinecite{dfgg3,sloane}, the meanders numbers $M_n$ up to $n=29$
are given.  With a recent computer and more tricks of programming, it is
perhaps possible to obtain $n=31$ or 32.  If the power of computers
continues to increase exponentially, the best we can expect with the full
enumeration without major improvement, is a linear growth in $n$, with a
rate of only one new size every 2 or 3 years.  We have the feeling that
this progress is too small to change significantly our understanding of the
meanders problem.

In this article, we will investigate larger $n$ with a Monte-Carlo
method increasing more slowly than an exponential.  But, this
stochastic method gives results with error bars and the interpretation
is more delicate.

%%%%%%%%%%%%%%%%%%%%%%%%%%%%%%%%
\section{The Monte-Carlo method}
%%%%%%%%%%%%%%%%%%%%%%%%%%%%%%%%

As explained in the previous section, the exponential growth of the
computations with an exact enumeration method limits the size of
meanders around $n\sim 30$.  To reach bigger $n$, it is then natural to
try to study this problem with a Monte-Carlo (or stochastic) method.

As the set of the $M_n$ meanders of a given size $n$ is too large to be
fully explored, the general idea is to randomly select a small subset.
Then, the measurements are done and averaged on the selected meanders.
It gives an {\sl estimator} of the exact (but unknown) result, with a
unknown error.  This error has two components : statistical
fluctuations and bias.

The statistical fluctuations can be reduced by independently repeating
the procedure many times.  Then, we obtain a histogram of the
estimator, with an average and a variance.  Under the hypothesis of
finite variance, the statistical fluctuations of the average can be
estimated by usual formulas of statistics.

The {\sl bias} is the difference between the exact result and the
mathematical expectation of the estimator.  If it can be exactly
calculated, we subtract it from the estimator.  But, in general, an
unknown part remains, which can not be reduced by a better statistics.
As explained below, by adjusting parameters of the simulation, the bias
can be reduced to become smaller than the statistical fluctuations.

In this section, we will first introduce the simplest algorithm, the
{\sl one-squirrel method}.  We will see that its statistical
fluctuations grow exponentially with $n$ and they are too big for $n
\approx 30$.  Then, we present an algorithm, the {\sl multi-squirrel
method}, for which the fluctuations increase less rapidly.

%%%%%%%%%%%%%%%%%%%%%%%%%%%%%%%%%%%%
\subsection{The one-squirrel method}
%%%%%%%%%%%%%%%%%%%%%%%%%%%%%%%%%%%%

The method is based on the recursion relation (see Fig.~\ref{recur}),
with which the set of meanders is organized as a tree (see
Fig.~\ref{arbre}).  The Monte-Carlo squirrel has the following
stochastic behavior.  It starts at the root of the tree (the meander of
size $n=1$).  It climbs into the tree.  At each level $n$, it stands on
a node and makes some measurements concerning the meander of size $n$,
represented by this node.  Then the squirrel goes to the level $n+1$ by
choosing at random one of the $b_n$ branches, starting from this node.
The squirrel stops at a prefixed level $n = n_{max}$.  This process
constitutes {\sl one} simulation.

The probability that the squirrel reaches a given meander of size $n$
is $ 1/\prod_{l<n} b_l$.  This probability law is not flat because the
sequence of $b_l$ depends on the visited nodes.  As seen in
Fig.~\ref{arbre}, for $n=5$, the two meanders on the left side have a
probability $1/8$ and the three on the right side have $1/12$.  So, to
correct this bias between meanders, the squirrel has a weight
  \begin{equation}
    q_n = \prod_{l=1}^{n-1} b_l,
  \end{equation}
which is calculated during its climbing.

By noting $\langle \cdot \rangle$ the mathematical expectation (over
all possible simulations), it is then obvious that $\langle q_n \rangle
= M_n$, because the sum runs over $M_n$ possible paths and the
contribution of a given path is $q_n$ with probability $1/q_n$.

More generally, for some quantity $Z$ (for example the winding number),
we wish to determine
  \begin{equation}
    {\cal Z} = \sum_{m=1}^{M_n} Z(m),
    \label {aa}
  \end{equation}
where $Z(m)$ is the value of $Z$ on the $m$-th meander of size $n$.
Each simulation gives, at level $n$, a measurement $Z(s)$ on the
meander $s$ reached by the squirrel and $\langle q_n \cdot Z(s) \rangle
= {\cal Z}$ which is the generalization of $\langle q_n \rangle = M_n$
obtained with $Z=1$.  Then
  \begin{equation}
    z = q_n\cdot Z(s)
  \end{equation}
is an {\sl unbiased} estimator of $\cal Z$ (i.e. $\langle z \rangle =
{\cal Z}$).

As usual in Monte-Carlo methods,  several simulations are made
independently and we hope that the average $\bar{z}$ of all the
measurements $z$ is close to ${\cal Z}$.  Unfortunately, this method
does not work in practice, because the weight $q_n$ is the product of
$b_l$.  Although the distribution of each $b_l$ is regular, the product
of many random variables is not self-averaging.

As the sum of $\ln(b_l)$ is self-averaging (i.e. the observed result is
closed to its mathematical expectation when $n$ goes to infinite), most
of the observed $q_n$ are not closed to $\langle q_n \rangle$ and
  \begin{equation}
    {\langle q_n \rangle \over q_n(\mbox{observed})} \sim
    \exp \sum_{l<n} (\ln \langle b_l \rangle - \langle \ln b_l \rangle)
  \end{equation}
increases like an exponential.  Then the averages with weight $q_n$ are
dominated by exponentially rare events and the statistical fluctuations
become large.  To keep the observed average close to the mathematical
expectation, the number of simulations must increase exponentially with
$n$ and fluctuations become too big for $n \sim 30$ or 35.  As the
difficulties increase exponentially with $n$ (as for exact
enumeration), is is useless to increase the power of the computer.  We
need a new algorithm.

%%%%%%%%%%%%%%%%%%%%%%%%%%%%%%%%%%%
\subsection{Multi-squirrels method}
%%%%%%%%%%%%%%%%%%%%%%%%%%%%%%%%%%%

\label{msm}

We generalize the one-squirrel method, but now with a population of $S$
squirrels, which reproduce and die ; $S$ is a fixed parameter during
all the simulations.  It is more simple to choose $S$ as $S = M_{n_0}$,
with at the starting point, a squirrel staying at each node of level
$n_0$ (meanders of size $n_0$).  In this work, $n_0 = 17$ and we use
the up-down symmetry to reduce the population to $S = M_{17}/2 =
1664094$ squirrels.

The population evolves from level $n$ to level $n+1$ by the following
process.  Each squirrel $i$ lives on a node $s_i$ on level $n$,
connected to $b_i$ nodes on level $n+1$.  It reproduces and has $b_i$
children and each child lives on each one of these $b_i$ nodes. The
total number of children $S'=\sum_{i=1}^S b_i$ is calculated.  The
ratio $B_n = S'/S$ is an estimate of $M_{n+1} / M_n$: the average of
the number of children per parent.  To prevent an exponential growth of
the population and of the needed computer memory and time, the total
population is keep constant by decimating the children : only $S$ among
the $S'$ children survive.  The choice is made at random with uniform
distribution.  Then the probability of surviving is $1/B_n$.  This
decimation is the single Monte-Carlo step of the algorithm.

This process is iterated up to reach a prefixed level $n = n_{max}$: it
gives {\sl one} simulation.  Many independent simulations are done and
averaged.

The particular case $S = 1$ gives the previous method with one
squirrel.  But, as for $S=1$, for every value of $S$, the probability
that a given meander is reached, is not uniform.  The nodes with small
number of ``brothers'' or ``cousins'' have always a small advantage.
But this bias becomes smaller when the population $S$ is large.  That
is the main improvement of this method.  The limit $S = \infty$
corresponds to the exact enumeration.

To correct the bias, each simulation has a weight 
  \begin{equation}
    q_n = \prod_{l=n_0}^{n-1} B_l
    \label{qn}
  \end{equation}
and the averages runs over all the simulations with their weight.  More
exactly, for some quantity $Z$, by keeping the notations of
Eq.~(\ref{aa}), one simulation with $S$ squirrels gives $S$
measurements $\{Z(s_i)\}$ for $i \in [1,S]$ with a weight $q_n$ and the
estimator
  \begin{equation}
    z = q_n \cdot \sum_{i=1}^S Z(s_i)
  \end{equation}
is unbiased, i.e.
  \begin{equation}
    \langle z \rangle = {\cal Z}.
    \label{moy}
  \end{equation}
We note that the case $Z=1$ gives $S \cdot \langle q_n \rangle = M_n$.

In order to prove Eq.~(\ref{moy}), we define the operator $\delta_m$
characterizing a given meander $m$ by $\delta_m(m') = 1$ when $m=m'$
and 0 otherwise.  Then every operator $Z$ can be split up into $Z =
\sum_m Z(m)\cdot \delta_m$.  As the expectation of a sum is always the
sum of expectations, we have to prove Eq.~(\ref{moy}) for the operators
$\delta_m$ only, which becomes $\langle  q_n \cdot \Delta_m \rangle =
1$, where $\Delta_m=1$ if the $m$-th meander is occupied by a squirrel
and 0 otherwise.  Let $p$ represent the parent of $m$ in the tree at
level $n-1$.  The probability that $\Delta_m=1$ (i.e. $m$ is occupied)
is the product of the  probability $\Delta_p $ that $p$ was occupied at
the level $n-1$ and the probability $1/B_{n-1}$ that its child $m$
survives after the decimation process $(n-1 \to n)$.

By using Eq.(\ref{qn}) and averaging on the random decimation $(n-1 \to
n)$ only,
  \begin{equation}
    \langle \Delta_m \; \prod_{l=n_0}^{n-1} B_l \rangle =
    \langle \Delta_p\;\prod_{l=n_0}^{n-2} B_l \rangle.
  \end{equation}
It is a recursion relation between a given meander of size $n$ and its
parent of size $n-1$.  By iterating, we go down to the ancestor at the
starting level $n_0$ for which $\Delta = 1$ and the empty product of
$B_l$ is 1.  It proves that $\langle  q_n \cdot \Delta_m \rangle = 1$
for every meander $m$ of size $n$, and Eq.~(\ref{moy}) is valid  for
every operator $Z$.

As usual, many simulations are done and the measurements are averaged
with their respective weight.  A priori, it seems that this method has
the same defect as the {\sl one}-squirrel method because the weight
$q_n$ is the product of many $B_l$, not self-averaging when $n$ becomes
large.  The ratio
  \begin{equation}
    {\langle q_n \rangle \over q_n(\mbox{observed})} \sim
    \exp \sum_{l=n_0}^{n-1}
    (\ln \langle B_l \rangle - \langle \ln B_l \rangle)
    \label{ratio}
  \end{equation}
between the mathematical expectation and the most frequently observed
$q_n$ increases like an exponential.  But, the main improvement of the
{\sl multi}-squirrel method is that the distribution of $B_l$ becomes
narrow when the population $S$ of squirrels is large. As $B_l = S'/S$,
the fluctuations of $B_l$ are of order $O(1/\sqrt S$) because the
number $S' = \sum_i b_i$ of children is the sum of $S$ random
variables.  A Taylor expansion of $\ln B_l$ shows that $\ln \langle B_l
\rangle - \langle \ln B_l \rangle = O(1/S)$

Then, with these simple arguments, we can hope that the ratio
(\ref{ratio}) grows like $1 + O(n/S)$ and that problems appear only
when $n$ becomes on the same order than $S$.  In our simulations, we
observe that the fluctuations grow with $n$ faster than this optimistic
prediction $O(n/S)$.  In fact, the $B_l$'s are not independent and the
exponential function accentuates all deviations.  Then we supervised
carefully the distribution of $q_n$.  When $n$ is small, we see a
regular bell-shaped curve.  But, when $n$ increases, the distribution
becomes asymmetric, with a long and irregular tail for the large
$q_n$.

For example, for $n=400$, with $S=1664094$ squirrels, the width
$\sigma$ of the distribution is only 12 \%, but we observed rare events
with a value of $q_n$ as big as three times the average.  However, in
this case, their contribution to the average and fluctuations is not
yet problematic.  But, if we let $n$ increase without control, rare
events will dominate and the results will become hazardous.

How to choose $n$ and $S$ ?  The na\"\i ve point of view is to take $n$
as bigger as possible.  But, to make $N_s$ independent simulations with
$S$ squirrels of size $n$, the need for computer memory is of order
$O(n\cdot S)$ and the need for computer time is of order
$O(n^2.S.N_s)$.  If $S$ is large enough, the fluctuations are Gaussian
and the error bars are of order $O(1/\sqrt{S.N_s})$.  As $n$ is always
limited, we will extrapolate to study the asymptotic behavior.  For
that, it is of no help to have large values of $n$ if the error bars
are too big.  So for a fixed computer time, we prefer accumulate good
statistics by limiting $n \leq 400$.  Finally for a fixed product
$S.N_s$, we prefer to take $S=1664094$ as bigger as permitted by the
memory computer to avoid the problem of rare but large fluctuations.

%%%%%%%%%%%%%%%%%%%%%%%%%%%%%%%%%%%%%%%%%%%%
\subsection{Bias for non-linear observables}
%%%%%%%%%%%%%%%%%%%%%%%%%%%%%%%%%%%%%%%%%%%%

\label{secbias}

In the previous section, we have seen how to obtain unbiased
Monte-Carlo estimates of the sum $\cal{Z}$ over all the meanders of
size $n$ of some quantity $Z$ (see Eq.~(\ref{aa}) and its notations).
However we are more interested by the {\sl average} ${\cal Z}/M_n$ over
all the meanders of size $n$.  For example, the average winding number
$w_n$ (see Eq.~\ref{wn}) is obtained when $Z$ counts the winding.  To
evaluate $R$ of Eq.~(\ref{rnng}), we can analyze $M_{n+1}/M_n$; in this
case, $Z$ counts the exterior arches.  More generally, we want to use
non-linear combinations of ${\cal Z}$ and $M_n$.

With our Monte-Carlo method, we have seen that one simulation gives a
measurement $z$ which is an {\sl unbiased} estimator: $\langle z
\rangle = {\cal Z}$.  With $N_s$ independent simulations, we call
$\bar{z}$ the usual average of the $N_s$ measurements $z$; its
fluctuations are $\sqrt N_s $ times smaller.  The bar over the symbols
distinguishes the average of observed values by Monte-Carlo method,
from the (unknown) mathematical expectation, marked with $\langle \dots
\rangle$.  The same work can be done with $q_n$ which is a unbiased
estimator of $M_n$.

We must be careful with the Monte-Carlo estimate of ${\cal Z}/M_n$.
For example, the average of the ratio $z/q_n$ gives bad results.  It is
better to compute the ratio of the averages $\bar{z} / \bar{q_n}$.
Indeed, with a Taylor expansion of $z$ and $q_n$ around their
mathematical expectations ${\cal Z}$ and $M_n$, the bias (defined as
the difference of the mathematical expectation $\langle \bar{z} /
\bar{q_n} \rangle$ and the target ${\cal Z}/M_n$)
  \begin{equation}
    \langle \bar{z} / \bar{q_n} \rangle - {\cal Z}/M_n = O(1/N_s).
    \label{aqn}
  \end{equation}
It can be neglected if it is smaller than the stochastic fluctuations.
Usually, in Monte-Carlo simulations, this problem disappears because
several millions of independent measurements are done.  But, in this
work, the situation is quite different. In fact, each simulation is a
complex process involving millions of squirrels, and the number $N_s$
of simulations is small.

Of course, we can not compute this bias exactly, otherwise we would
have already subtract it from measurements.  But we can estimate it by
the following process.  The set of simulations is divided into
$N_s/2^p$ subsets, with $2^p$ simulations each.  In each subset,
$\bar{z} / \bar{q_n}$ is computed.  We obtained $N_s/2^p$ independent
values, one for each subset; by usual formulae of statistics, we
compute their average $E^{(p)}$ and the error bars.  This work is done
for all integer $p$ between $p=0$ (each subset contains only one
simulation) and $p=\log_2 N_s$ (only one set with all the $N_s$
simulations).

Which value of $p$ is the best ?  Following Eq.~(\ref{aqn}), the bias
of $E^{(p)}$ is expected to decrease like $1/2^p$.  For small values of
$p$, we observe really a dependency of $E^{(p)}$ on $p$: the bias is
visible.  But, for $p>5$, variations become smaller than statistical
error bars:  the size $2^p$ of subsets is sufficiently large to neglect
the bias.  But, if $p$ is close to its maximum, the number of subsets
becomes very small and the error bars are not properly estimated.  As
in our work $N_s = 8192 = 2^{13}$, we finally keep $p=7$: the estimator
$E^{(7)}$ is computed with 64 independent subsets of 128 simulations
each.  This method was used for all quantities presented below.  It is
valid, not only for ratios like ${\cal Z}/M_n$, but also for non-linear
functions like $\ln(M_{n+1}/M_n)$.  It could also be possible to use
more complex estimators.  For example, the combination $2E^{(p)} -
E^{(p-1)}$ cancels the order $1/2^p$ of the bias.

%%%%%%%%%%%%%%%%%
\section{Results}
%%%%%%%%%%%%%%%%%

In this section, we describe the results obtained by our Monte-Carlo
multi-squirrels method.  After several tests, we used a population of
$S=1664094 (=M_{17}/2)$ squirrels for meanders with size up to
$n=400$.  To do $N_s=8192$ independent simulations, we have used during
8 days a parallel computer (the Cray T3E of the Cea-Grenoble) with 128
processors (Dec-alpha at 375 MHz) and 13 Gigabytes of total memory,
equivalent to 24000 hours of workstation cpu time.

We have verified that the results are stable when $S$ (the population)
increases.  More exactly, the tests with smaller $S$ have larger error
bars, but are compatible with results and error bars obtained with the
largest $S$.  As explained in the Sect.~\ref{msm}, we have carefully
checked that $S=1664094$ is sufficiently large to explore sizes of
meander up to $n=400$.

%%%%%%%%%%%%%%%%%%%%%%%%
\subsection{Enumeration}
%%%%%%%%%%%%%%%%%%%%%%%%

We want to measure $R$ and $\gamma$, which describe the asymptotic
behavior of the number of meanders $M_n \sim c \; R^n/n^\gamma$ for
large $n$.  The entropy $\ln R$ can be estimated by $\ln
(M_n/M_{n-1})$.  But it appears that the sub-sequences $M_{2n}$ and
$M_{2n+1}$ have an alternating sub-leading correction.  We have
estimated it to be $u (-1)^n / (n\ \ln n)$ with $u = 0.5(1)$.  This
alternating effect is dramatically amplified by the ratio
$M_n/M_{n-1}$.  So it is better to consider
  \begin{equation}
    L_n = {1 \over 2} \ln (M_n/M_{n-2}),
  \end{equation}
with a jump from $n-2$ to $n$.  But even with this precaution, the
reader can still see on the following figures a small parity effect.
To estimate $L_n$, we have used the procedure described in
Sect.~\ref{secbias}.

\begin{figure}
  \centering\leavevmode
  \epsfxsize=8cm
  \epsfbox{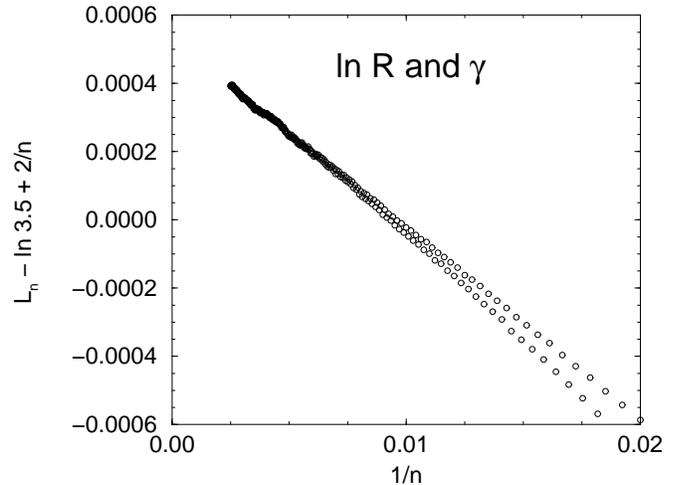}
  \caption{
    $\ln R$ and $\gamma$~: plot of the Monte-Carlo estimate of $L_n -
    \ln 3.5 + 2/n$, for n between 50 and 400, versus $1/n$. The limit
    when $x$ goes to 0 is $\ln (R/3.5)$, and the (negative) slope is
    $2-\gamma$.  The error bars are not drawn~; their maximum is
    $10^{-5}$, then they are smaller than the symbols.  A parity
    effect, between the odd and even $n$, is visible.  A linear
    extrapolation gives $R = 3.5019(2)$ and $\gamma = 2.056(10)$.
  }
  \label{lnr}
\end{figure}

\begin{figure}
  \centering\leavevmode
  \epsfxsize=8cm
  \epsfbox{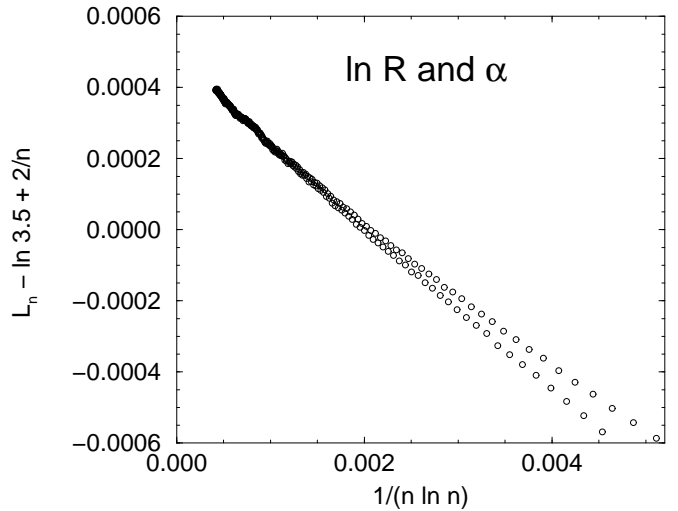}
  \caption{
    $\ln R$ and $\alpha$ under the hypothesis $\gamma=2$~: the same
    plot as Fig.\ref{lnr} but the $x$-axis is $1/(n\ln n)$. The
    (negative) slope is $-\alpha$.  A linear extrapolation gives $R =
    3.5017(2)$ and $\alpha = 0.25(5)$.
  }
  \label{lnrlog}
\end{figure}

As we expect $L_n \sim \ln R - \gamma / n$ for large $n$, by plotting
$y=L_n$ versus $x = 1/n$, we can estimate $\ln R$ (limit when $x$ goes
to 0) and the exponent $\gamma$ (asymptotic slope).  In Fig.~\ref{lnr},
we have plotted the Monte-Carlo estimate of $L_n - \ln 3.5 + 2/n$
versus $1/n$ for $n$ between 50 and 400.  We have arbitrarily
subtracted the {\sl linear} function $y = \ln 3.5 - 2x$, to reduce the
amplitude of $y$; we obtain a figure where the small quantities
$2-\gamma$ (remaining slope), $\ln(R/3.5)$ (limit when $x$ goes to 0)
and curvature (deviation to the expected linear behavior) are more
visible.  The curvature remains small and a linear extrapolation gives
a limit between $0.0005$ and $0.0006$ with an estimated slope
$0.056(10)$.  Then
  \begin{equation}
    R = 3.5019(2) \;\;\;\; \mbox{ and } \;\;\;\; \gamma = 2.056(10).
    \label{b1}
  \end{equation}

With the assumption that the asymptotic behavior is $M_n \sim c \;
R^n/n^\gamma$, the conjecture $\gamma = 2$ is incompatible with these
simulations.  But, we can try another asymptotic shape, for example
  \begin{equation}
    M_n \sim c \; {R^n \over n^\gamma}\;{1 \over \ln^\alpha n},
    \label{alpha}
  \end{equation}
by introducing a new exponent $\alpha$.  In Fig.~\ref{lnrlog}, we have
plotted $L_n - \ln 3.5 + 2/n$ (as in Fig.~\ref{lnr}), but now versus
$1/(n \ln n)$.  With this transformation of the $x$-axis, a linear
behavior corresponds to $\gamma = 2$ and the slope measures $-\alpha$.
A linear extrapolation gives
  \begin{equation}
    R = 3.5017(2), \;\;\;\; \gamma = 2 \;\;
       \mbox{ and } \;\;\;\; \alpha = 0.25(5).
    \label{b2}
  \end{equation}
with $\alpha$ compatible with the simple fraction $1/4$.

How shall we choose between both results Eq.~(\ref{b1}) or
Eq.~(\ref{b2}) ?  We notice that the quality of the alignment of points
is the same in Fig.~\ref{lnr} and Fig.~\ref{lnrlog}.  In fact, it is
very difficult to distinguish between a logarithmic law and a power law
with such a small exponent, with statistical error bars and when the
amplitude of $\ln n$ is small.  In fact, for $n$ close to $n'$, the
term $\alpha/(n \ln n)$ looks like $\alpha'/n$ with $\alpha' / \alpha =
1/\ln n' + 1/\ln^2 n'$.  Then, for any choice of $\alpha$ not too
large, $\gamma = 2.056 - 0.22 \alpha$ gives a class of acceptable
behaviors.

We have tried to use more sophisticated extrapolation methods.  For
example, with a fixed jump $i$, $(n L_n - (n-i)L_{n-i})/i$ gives
theoretically the same limit $\ln R$ but by removing the term
$\gamma/n$.  Thus $n^2 (L_n - L_{n-i})/i$ gives a direct estimate of
$\gamma$.  Unfortunately these kinds of derivative amplify statistical
errors, and the results are compatible but less precise than the
previous estimates.

So, with our numerical simulations, we cannot say if a logarithmic
factor is present or not.  However, $R = 3.5018(3)$, and we can exclude
$R=3.5$ and the conjecture $\gamma=2$ without logarithmic factor
($\alpha=0$).

%%%%%%%%%%%%%%%%%%%%%%%%
\subsection{Winding}
%%%%%%%%%%%%%%%%%%%%%%%%

\begin{figure}
  \centering\leavevmode
  \epsfxsize=8.5cm
  \epsfbox{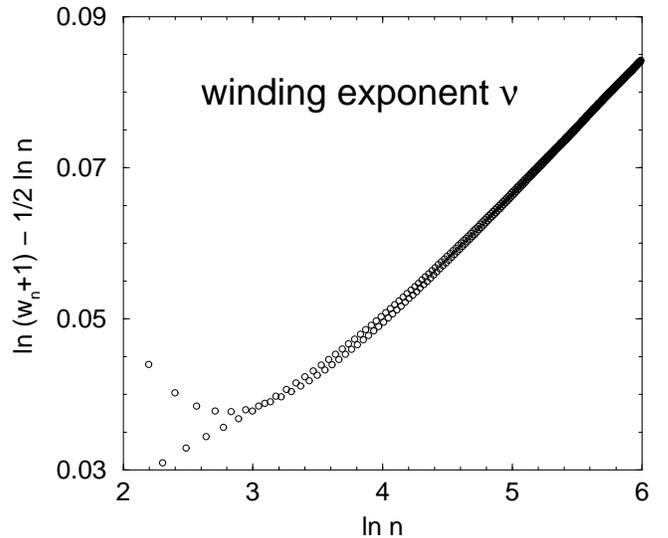}
  \caption{
    Winding exponent $\nu$ : plot of the Monte-Carlo results of
    $\ln(w_n+1) - 1/2\ \ln n$, for $n$ between 8 and 400, versus $\ln
    n$.  The slope is $0.018$; it is a measurement of $\nu - 1/2$. The
    error bars are not drawn; their maximum is $3.10^{-4}$, then they
    are smaller than the symbols.
  }
  \label{fnu}
\end{figure}

We will present the Monte-Carlo results of the exponent $\nu$, which
describe the asymptotic behavior of the average winding number $ w_n
\sim n^{\nu}$.  To avoid problem with bias, we have used the procedure
described in Sect.~\ref{secbias}.  By plotting $\ln(w_n)$ versus
$\ln(n)$, the asymptotic slope will be a measurement of $\nu$.  In
Fig.~\ref{fnu}, we have plotted $y = \ln(w_n+1) - {1/2} \ln n$ versus
$x = \ln n$.  We have arbitrarily considered $\ln(w_n + 1)$ and not
$\ln(w_n)$ because $w_n+1$ is less sensitive than $w_n$ to the finite
size effects~\cite{dfgg3}.  As the main question is to know whether
$\nu = 1/2$ or not, we have arbitrarily subtracted the linear function
$y = x/2$.  Then the variation of $y$ is reduced and we obtain a figure
where $\nu - 1/2$ (residual slope) and the curvature are more visible.
We see that the curvature is small and a linear extrapolation gives
$\nu = 0.518$.  As it is difficult to estimate the errors with the data
of Fig.~\ref{fnu}, we have also tried more sophisticated quantities
like
  \begin{equation}
    G_i(n) = {n \over i} \ln \left( {w_n+1 \over w_{n-i}+1} \right)
  \end{equation}
which are discrete derivatives of $\ln (w_n+1)$ with step $i$.  They
give a direct value for $\nu$, but unfortunately the statistical
fluctuations are amplified by this differentiation and the uncertainty
over $\nu$ is of order $0.002$.

We have seen that, for the exponent $\gamma$, a behavior with
logarithmic correction is not excluded by the Monte-Carlo data.  So we
have tried to fit the winding number with
  \begin{equation}
    w_n \sim n^{1/2} \ln^{\alpha} n.
  \end{equation}
With this hypothesis, a plot of $y = \ln(w_n+1) - {1/2} \ln n$, as in
Fig.~\ref{fnu}, but now versus $x = \ln \ln n$ would give a straight
line with slope $\alpha$.  But the curvature is much stronger than that
of Fig.~\ref{fnu}.  So we dismiss this hypothesis and conclude that
  \begin{equation}
    \nu = 0.518(2)
    \label{vnu}
  \end{equation}

With the assumption that the asymptotic behavior is a simple power-law,
the Brownian value $\nu = 1/2$ is incompatible with these simulations.

%%%%%%%%%%%%%%%%%%%%%%%%%%%%%%%%%%%%%%%%%%%%%%%%%%%%%%%
\subsection{Probability distribution of winding number}
%%%%%%%%%%%%%%%%%%%%%%%%%%%%%%%%%%%%%%%%%%%%%%%%%%%%%%%

\begin{figure}
  \centering\leavevmode
  \epsfxsize=8.5cm
  \epsfbox{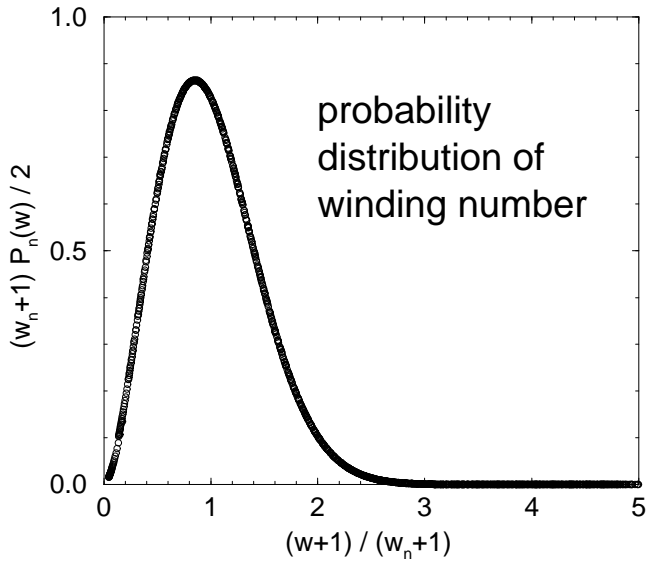}
  \caption{
    Plot of $(w_n+1) P_n(w)/2$ versus $(w+1)/(w_n+1)$ for $n=$30, 40,
    50 \dots 400. The points accumulate on a scaling function $f$.
    The error bars are not drawn; their maximum is $6.10^{-4}$.
  }
  \label{pnw}
\end{figure}

We define the probability distribution $P_n(w)$ of winding number as
the fraction of the meanders of size $n$ with $w$ windings.  We
expect~\cite{dfgg3} the asymptotic scaling behavior
  \begin{equation}
    P_n(w) \approx {2 \over w_n+1} \;\; f \left({w+1 \over w_n+1}\right)
  \end{equation}
with a scaling function $f(x)$, where $w_n$ is the average winding
number. With the factor 2, the integral of $f$ is normalized to 1
because $w$ and $n$ are integers with the same parity.  In
Fig.~\ref{pnw}, we plot $y = (w_n+1) P_n(w) / 2$ versus $x =
(w+1)/(w_n+1)$ for different values of $n$.  To define the scaling
variable $x$, we prefer to take $w+1$ instead of $w$ to reduce finite
size effects.

We see that the points accumulate on a smooth curve, which represents
the scaling function $f(x)$.  By analogy with the end-to-end
distribution for polymers, we expect~\cite{dfgg3} a power law behavior,
$f(x) \sim x^{\theta}$, for small $x$, and a behavior $f(x) \sim
\exp(-\mbox{const.}x^{\delta})$, for large $x$.  Our data give the
estimates
  \begin{equation}
    \theta = 1.7(1) \;\;\; \mbox{and}\;\;\; \delta = 2.3(1).
    \label{vtheta}
  \end{equation}
To obtain a better precision, it would require to have larger values of
$n$.

%%%%%%%%%%%%%%%%%%%%%%%%%%%%
\subsection{Height and area}
%%%%%%%%%%%%%%%%%%%%%%%%%%%%

\label{haa}

We are interested by the asymptotic behavior, for large size $n$, of
the average area $A_n$ (see Eq.~\ref{av.area}) and average height
$h_n(i)$ (see Eq.~\ref{av.height}).

\begin{figure}
  \centering\leavevmode
  \epsfxsize=8.5cm
  \epsfbox{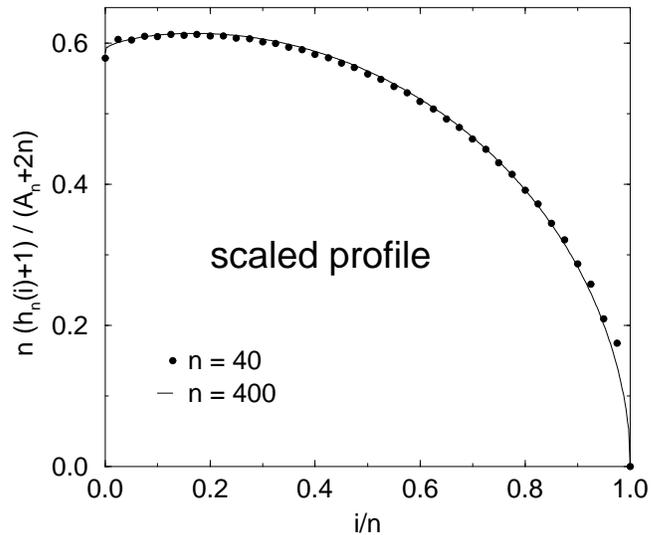}
  \caption{
    Plot of the scaled height $n (h_n(i)+1) / (A_n + 2n)$ versus the
    scaled coordinate $x=i/n$.  Only two set of data ($n=40$ and 400)
    are shown in order not to overload the figure.  The points
    accumulate on a scaling function $\rho(x)$.  We can see small
    deviations for $x=0$ and $x=1$, which are analyzed in the text.
  }
  \label{profile}
\end{figure}

The label $i$ is the ``horizontal'' coordinate and varies between $-n$
and $n$.  So, we introduce the scaled variable $x = i/n$, with $-1 \leq
x \leq 1$.  We expect~\cite{dfgg3} that
  \begin{equation}
    h_n(n x) \sim {A_n \over n} \rho(x)
    \label{rho}
  \end{equation}
for large $n$ by fixing $x$, where $\rho(x)$ is a scaling function with
integral normalized to 1.  In Fig.~\ref{profile}, we have plotted $y =
n \ (h_n(i)+1) / (A_n + 2n)$ versus $x = i/n$ for various size $n$.
Only the positive $i$ are shown because $h_n(i)$ is symmetric after
summing over all the meanders.  More precisely, we have plotted the
average of $\{h_n(i)+h_n(-i)\}/2$ over the Monte-Carlo samples of
meanders, which is equivalent to the average of $h_n(i)$ over these
samples, plus those obtained by the left-right symmetry ($i \rightarrow
-i$).  As in the previous figures, we prefer to take $(h_n+1)$ and
$(A_n+2n)$ to reduce finite size effects.

We see that, for $0<x<1$, the points accumulate on a smooth curve,
which represents the scaling function $\rho(x)$.  This shape is not a
half-circle, as it would be for a random-walk on a semi-infinite
line~\cite{dfgg3}.

As $h_n(0)$ is the winding number $w_n$ for the particular case $x=0$,
Eq.~(\ref{rho}) can be valid only if $h_n(nx)$ scales as $n^{\nu}$ with
the same exponent $\nu$ for all $x$.  Consequently, the area $A_n$
scales as $n^{\nu+1}$.  In a previous section, we have numerically
determined $\nu = 0.518(2)$ by extrapolation of $w_n=h_n(0)$.  The same
work with $h_n(n/2)$ and $A_n$ gives $\nu = 0.517$ for both, compatible
with the previous estimates, but less precise because the finite size
effects are stronger.

In Fig.~\ref{profile}, small deviations appear between the curves for
$n$=40 and 400, at the boundary ($x \simeq \pm 1$) and in the middle of
the meander ($x \simeq 0$).

\begin{figure}
  \centering\leavevmode
  \epsfxsize=8.5cm
  \epsfbox{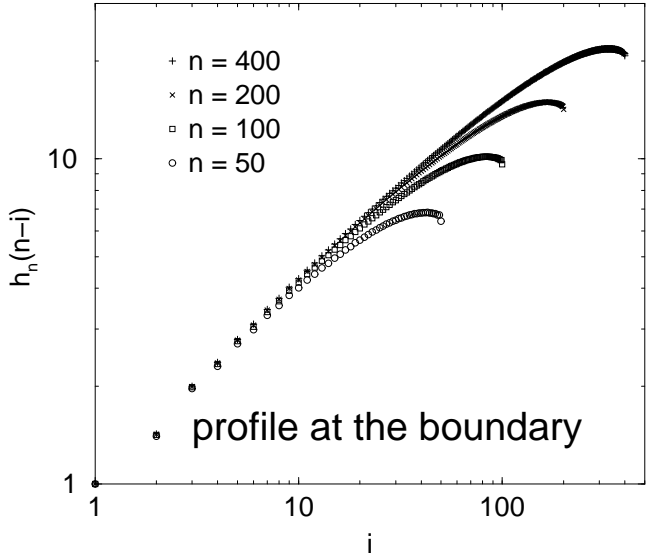}
  \caption{
    Log-log plot of the height $h_n(n-i)$ versus the distance to the
    boundary $i$ for various sizes $n$.  The points accumulate on a
    limiting curve, which can be fitted by $i^{\phi}$ with $\phi =
    0.64(2)$.
  }
  \label{bord}
\end{figure}

In order to understand why finite-size effects are important near the
boundary, Fig.\ref{bord} is a plot of $h_n(n-i)$ versus $i$, with
various $n$.  Clearly, when $n$ is large, curves accumulate on a
limiting curve
  \begin{equation}
    \tilde{h}(i) = \lim_{n \rightarrow \infty} h_n(n-i).
  \end{equation}

As this function can be fitted by a straight line on this ``log-log''
plot, we define a new exponent $\phi$ by
  \begin{equation}
    \tilde{h}(i) \sim i^{\phi} \mbox{\ \ with \ \ } \phi = 0.64(2)
    \label{htilde}
  \end{equation}
when $i$ is large.  As by construction, $h_n(n) = 0$ and $h_n(n-1) = 1$
for all $n$, $\tilde{h}(0)=0$ and $\tilde{h}(1)=1$ satisfy exactly
Eq.~(\ref{htilde}).

If we set $i = ny$, by using Eq.~(\ref{htilde}) valid when $i$ is
fixed, we obtain $h_n(n-i) \sim (ny)^{\phi}$.  On the other hand, by
using Eq.~(\ref{rho}) which is valid when $y$ is fixed, we obtain now $
n^{\nu} \rho(1-y)$.  As the exponents $\nu$ and $\phi$ differs, these
two regimes are incompatible and the behavior of $h_n(n-i)$ depends on
the order in which $n$ and $i$ go to infinity.  In particular, the
domain of validity of Eq.~(\ref{htilde}) is reduced to the single point
$x=1$ for Eq.~(\ref{rho}).  That explains why the rightmost dots ($i
\sim n$) in Fig.~\ref{profile} for the small size are not superimposed
on the curve obtained for the large size.

Here, the exponent $\phi$ is the {\sl surface} critical exponent, while
$\nu$ is the {\sl bulk} critical exponent.  Near the boundary, $h(i)$
is small and the condition $h(i) \geq 0$ limits appreciably the
fluctuations toward the bottom.  This effect is so strong that the
exponent is changed and $\phi > \nu$.  This situation is reminiscent of
other critical phenomena~\cite{binder8}, like the self-avoiding walk
near a surface.  Our result is to be contrasted with the case of a
random walk on a semi-infinite line for which the surface exponent
keeps its Brownian value $1/2$.

\begin{figure}
  \centering\leavevmode
  \epsfxsize=8.5cm
  \epsfbox{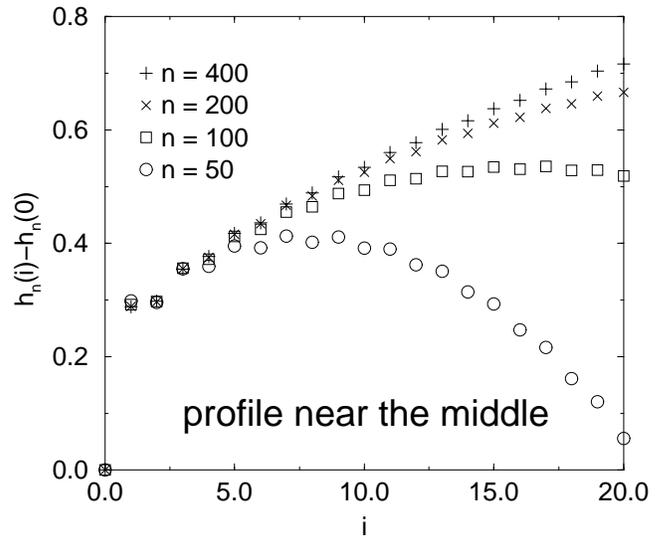}
  \caption{
    Plot of the $h_n(i) - h_n(0)$ versus $i$ for small $i$, near the
    middle of meander, for various size $n$.  By symmetry, only the
    positive $i$ are drawn.  As the heights $h_n(i)$ are shifted by
    $h_n(0)$ and {\sl not} scaled, we observe effects smaller than the
    asymptotic behavior $n^{\nu} \rho(0)$.  The points accumulate on a
    limiting curve $\hat{h}(i)$.
  }
  \label{cusp}
\end{figure}

Finite-size effects observable near the middle of meander ($x \sim 0$)
can be explained with Fig.~\ref{cusp}, which is a plot of
$h_n(i)-h_n(0)$ versus $i$ for small $i$ with various $n$.  Clearly,
when $n$ is large, curves accumulate on a limiting curve
  \begin{equation}
    \hat{h}(i)= \lim_{n \rightarrow \infty} ( h_n(i) - h_n(0) ).
  \end{equation}

If we make the hypothesis that the behavior of $\hat{h}(i)$ is
compatible with Eq.~(\ref{rho}) by inverting the limits $n$ large and
$x$ small, the consequences would be that $\rho(x)$ has a cusp at $x=0$
with a infinite derivative $ \rho(x) \sim \rho(0) + x^{\nu} $ when $x$
is small, and $ \hat{h}(i) \sim i^{\nu} $.  But this power law behavior
of $\hat{h}(i)$ is not observed.  Then the asymptotic behavior of
$h_n(i)$ with $i$ fixed and $n$ large is given by Eq.~(\ref{rho}),
i.e.  $n^{\nu} \rho(0)$, plus finite corrections of order
$\hat{h}(i)$.

This cusp is an another boundary effect since the point $i=0$ is the
source of the river.  Let us consider three consecutive heights
$\{h(i-1,m), h(i,m), h(i+1,m)\}$ for a given meander $m$.  By
definition, $h(i+1,m) = h(i,m) \pm 1$.  Then, for a generic $i$, the
couple of its neighbors can have 4 respective values $\{h(i-1),h(i+1)\}
= \{h(i)\pm 1,h(i)\pm 1\}$.  But, for the special case $i=0$, the
situation $h(-1) = h(1) = h(0)-1$ happens only if a single arch
connects the first bridge (near the source) to itself, by drawing a
little circle around the source, without visiting the other bridges.
This is forbidden if we insist in having only one connected component.
In other word, the neighborhood of the source limits the fluctuations
of $h(0)$ toward the top.  In particular, for every meander, $h(0) \leq
\{h(-1)+h(1)\}/2$.  That explains why, on average, $h(0) < h(1)$.

To understand why $h(1) < h(2)$, it is mandatory to consider more
complex forbidden situations for $\{ h(-2)$, $h(-1)$, $h(0)$,
$h(1)$, $h(2) \}$.  For example, a circle connecting bridges 1 and 2
with an upper and a lower arch, with $h(\pm 1) = h(\pm 2) + 1$, is
forbidden.

More generally, for a given $i$, the presence of the source forbids the
systems of arches connecting the bridges $\{1,2,\ldots,i\}$ with a
closed road without visiting the other bridges $\{i+1, \ldots\}$.
Qualitatively, it is a repulsive ``force'' which favors the connection
of bridge $i$ with bridges $j>i$ and gives a concave shape for small
$i$.

As the forbidden situations are more and more complex when $i$ becomes
large, their statistical effects decrease and this repulsive force has
a finite range.  In the end, the summation over all forbidden
situations gives finally this cusp with a finite amplitude described by
$\hat{h}(i)$.

%%%%%%%%%%%%%%%%%%%%
\section{Conclusion}
%%%%%%%%%%%%%%%%%%%%

In this paper, we have presented a Monte-Carlo method to investigate a
phase space described by a deterministic but irregular tree (i.e. the
number of branches at each node is not fixed).  With a na\"{\i}ve
random climbing on the tree (the {\em one}-squirrel method), the
probability of a path depends on the number of branches encounter at
each node.  For the meanders problem, the ratio between extremal
weights increases exponentially with the size~:  consequently the most
part of the computer time is devoted to generate configurations with
small weight, and only a exponentially small number of configurations
with high weight contribute efficiently to the average.

With the {\em multi}-squirrel method, the distribution becomes almost
flat~: the bias, i.e. the ratio between extremal weights increases very
slowly and never exceeds 3 in our simulations.  Moreover, this bias is
exactly known during the simulation, then it can be corrected to
average over all meanders with a uniform distribution.  As usual with
Monte-Carlo simulations, results suffer from statistical fluctuations
which decrease, in the best case, like the square root of the computer
time.

After a simulation on a parallel computer with 3 years of cpu time
in single processor units,
results with small errors bars have been obtained for meanders up to
size $n=400$.  Under some hypothesis inspired by the analogy with
random walks problems, large $n$ extrapolation can be done for the
enumeration (see Eq.~\ref{b1} and Fig.~\ref{lnr}), for the distribution
(see Eq.~\ref{vtheta} and Fig.~\ref{pnw}) and the average of the
winding number (see Eq.~\ref{vnu} and Fig.~\ref{fnu}) and the shape
(see Section~\ref{haa}) of meanders.

From a Monte-Carlo point of view, the proposed algorithm can be used,
in principal, for any combinatorial problem described by a tree~: the
essential ingredient is to know, for a given node, the number of
branches.  However algorithms are rare in this kind of problems and
better variants or other algorithms could be without doubt invented.

About the meanders, with these estimates, it appears that the critical
exponent are not simple fractions as $1/2$ or $7/2$, as conjectured by
previous studies~\cite{lando_zvonkin}.  Of course, Monte-Carlo
simulations cannot determine the exact values, but can confirm or
invalidate analytical proposals, while waiting for a rigorous
solution.

\acknowledgements

We thank L. Colombet, P. Di Francesco, E. Guitter and R. Napoleone for
stimulating discussions, critical reading of the manuscript and help
for an efficient parallelization of the computer program.

%%%%%%%%%%%%%%%%%%

\end{document}